\documentclass[english]{article}
\usepackage{graphicx}
\def\be{\begin{equation}}
\def\ee{\end{equation}}
\def\bea{\begin{eqnarray}}
\def\eea{\end{eqnarray}}
\def\bma{\begin{mathletters}}
\def\ema{\end{mathletters}}
\def\bi{\begin{itemize}}
\def\ei{\end{itemize}}

\def\C{\hbox{$\mit I$\kern-.7em$\mit C$}}

\newcommand{\ket}[1]{ | \, #1  \rangle}

\begin{document}
\begin{center}
{\large{Comment on ``Local Copying of $d\times d$-dimensional
Partially Entangled Pure States"}}\\
\vspace{0.3in}
Ramij Rahaman  \\
Selmer Center, Department of Informatics, University of Bergen, Bergen, Norway \\
\emph{Ramij.Rahaman@ii.uib.no}       
\end{center}

\vspace{0.3in}
\begin{abstract}
Recently, Li. \emph{et. al.} [Int. J. Theor. Phys., 48, 2777
(2009)] derived a necessary and sufficient condition for LOCC
cloning of a set of bipartite orthogonal partially but equally
entangled state. We demonstrates that, the result is based on a
wrong observation regarding a set of non-maximally entangled
states with equal entanglement. We also provide a simple example
in favor of our comment.

\end{abstract}

\section{Introduction}
In a recent publication by Li and Shen \cite{li} derive the
necessary and sufficient condition for local
copying\footnote{cloning under local operation and classical
communication (LOCC)} of a set of bipartite orthogonal partially
but equally entangled (BOPEE) states. To established the result
they claim to have derived a relation to hold for a set of BOPEE
states. Let $\{\ket{\Psi_j}\}_{j=1}^n$ be a set of BOPEE states.
According to their claim the states can be expressed as
$\ket{\Psi_j}=(U^1_j \otimes I^2) \ket{\Psi_1}$ ($U_j$'s are the
unitary operator acting on first system and $I$ is the identity
operator acting on the second system). But this is not true in
general. Though the results of \cite{li} have no contradiction
with the necessary condition given in \cite{choudhary,kay} for
Local copying of a set of BOPEE states in various cases. In
particular, for maximally entangled states, the above relation is
true \cite{ghosh1,anselmi,owari}, whereas this may not hold even
for a pair of BOPEE states.\\

To show this we consider the following pair of states\\
$\ket{\Psi_1}=a\ket{00}+b\ket{11}$ and
$\ket{\Psi_2}=b^*\ket{00}-a^*\ket{11}$, where $`*`$ indicate the
complex conjugate and $|a|^2+|b|^2=1;~|a|\neq|b|;~0<|a|,|b|<1$.
Here $\{\ket{\Psi_j}\}^{2}_{j=1}$ is a set of two BOPEE states.
Now we show that $\ket{\Psi_2}$ does not satisfy the relation
$\ket{\Psi_2}=(U^1_2 \otimes I^2) \ket{\Psi_1}$, for any unitary
$U_2^1$ acting on first Hilbert
space $\mathcal{H}^1$.\\
If possible, we assume that, there exists a $2\times 2$ unitary
operator
such that $\ket{\Psi_2}=(U\otimes I) \ket{\Psi_1}$ hold.\\

The general form of a $2\times 2$ unitary matrix is $U=
\left(\begin{array}{cc}
        \alpha & \lambda \beta \\
        -\beta^*& \lambda \alpha^*
        \end{array}
  \right) $, where $\alpha, \beta , \lambda $ are complex and
  $|\alpha|^2+|\beta|^2=1=|\lambda|$\\
  If $\ket{\Psi_2}=(U\otimes I) \ket{\Psi_1}$ holds, then from
  simple algebra we have the following equations.\\

  \begin{eqnarray}
\label{re1} a \alpha &=& b^*\\
\label{re2}b \lambda \beta &=& 0 \\
\label{re3}-a \beta^* &=& 0\\
\label{re4}b \lambda \alpha^*&=& -a^*
\end{eqnarray}

From equation (\ref{re2}) and equation (\ref{re3}) we have,
\begin{eqnarray}
 \beta =0 \mbox{ ~~~~   (since, $a\neq 0 \neq b$ and $|\lambda| =1$) }
\end{eqnarray}
Therefore, $|\alpha|=1$.\\
Now equation (\ref{re1}) and equation (\ref{re4}) have solution
only if
$|a|=|b|$, as, $|\alpha|=1=|\lambda|$.\\
$|a|=|b|$ imply that both $\ket{\Psi_1}~ \& ~\ket{\Psi_2}$ are
maximally entangled states.\\
Therefore, for non-maximal state, equ. (\ref{re1}-\ref{re4}) are
inconsistent, which imply that $\ket{\Psi_2}$ can't be expressed
as $\ket{\Psi_2}= (U\otimes I)
\ket{\Psi_1} $, for any $2 \times 2 $ unitary operator $U$.\\

Let us now point out the wrong step in their \cite{li} derivation,
which
led them to this wrong result.\\

Let $\{ \ket{e_i}\}_{i=1}^d$ be an orthogonal basis for single
particle Hilbert space $\mathcal{H}$ of dimension d.
$\ket{\Phi_{1}}=\frac{1}{\sqrt{d}}\sum_{i=1}^d
{\ket{e_i}\ket{e_i}}$ is a maximally entangled state in
$\mathcal{H}^{\otimes 2}$ and $\ket{\Psi_{j}}=\sum_{i=1}^d
{\alpha_i^j \ket{e_i}\ket{e_i}}$ are the non-maximally entangled
states in $\mathcal{H}^{\otimes 2}$, for $j=1,2,....,n$, with
$\sum_{i=1}^d {|\alpha_i^j|^2}=1$ .\\
 Now it is true that for
every pure bipartite non-maximally entangled state $\ket{\Psi_1}$,
there exist, a POVM's\footnote{Positive Operator-Valued Measure}
$\mathcal{M}^1$ acting on the first Hilbert space $\mathcal{H}^1$,
such that

\begin{equation}
\label{nonmax} \ket{\Psi_1}=(\mathcal{M}^1 \otimes I^2)
\ket{\Phi_{1}}
\end{equation}

Let $\{\ket{\Psi_j}\}^{n}_{j=1}$ be a set of BOPEE states. Then
the following relation holds

\begin{equation}
\label{nonmax1} \ket{\Psi_j}=(V_j^1 \otimes W_j^2) \ket{\Psi_{1}}
\end{equation}
for all $j(=1,2,...,n)$, $V_j^1$ and $W_j^2$ being unitary on
$\mathcal{H}^1$
 and $\mathcal{H}^2$ respectively.

In particular $V_1^1=I^1$ and $W_1^2=I^2$. From equation
(\ref{nonmax}) and equation (\ref{nonmax1}), we have

\begin{equation}
\label{nonmax2} \ket{\Psi_j}=(V_j^1 \otimes W_j^2)(\mathcal{M}^1
\otimes I^2) \ket{\Phi_{1}}
\end{equation}

Li et.al. has rewritten the equation (\ref{nonmax2}) as
\begin{equation}
\label{nonmax3}\ket{\Psi_j}=(\mathcal{M}^1 \otimes I^2)(V_j^1
\otimes W_j^2) \ket{\Phi_{1}}
\end{equation}
 from which the desired relation \cite{relation}
\begin{equation}
\label{nonmax4} \ket{\Psi_j}=( I^1\otimes U_j^2)\ket{\Psi_{1}}
\end{equation}
follows. But the problem is that equation (\ref{nonmax3}) may not
follow from eqation (\ref{nonmax2}) as $(\mathcal{M}^1 \otimes
I^2)$ and $(V_j^1 \otimes W_j^2)$ may not commute in general.

  \section{Conclusion}

 In conclusion, we have pointed out an error in the derivation of a result
 needed to prove a theorem on local cloning of orthogonal
 entangled states given in \cite{li}. The interesting problem of finding the necessary and sufficient
condition
 for local copying of arbitrary set of bipartite orthogonal partially but
equally entangled states still remains open.  \\

\section{Acknowledgments}
 The research has been supported by Norwegian Research Council. The author is grateful to G. Kar and S. Ghosh for fruitful discussions.

\end{document}